\documentclass[super,prb,
twocolumn,reprint,
groupedaddress,superscriptaddress,
amsfonts,amssymb,amsmath,floatfix,
showkeys,showpacs,a4paper]{revtex4-1}

\usepackage{CJK}

\usepackage{graphicx}
\usepackage[english]{babel}
\usepackage{hyperref}
\usepackage{upgreek}
\usepackage{natbib}
\usepackage{lineno}
\usepackage{csquotes}
\usepackage{color}
\usepackage{mathpazo}

\begin{document}

\title{Role of intermediate 4$f$ states in tuning the band structure of high entropy oxides}

\author{Abhishek Sarkar}\email{abhishek.sarkar@kit.edu}
\affiliation{Joint Research Laboratory Nanomaterials – Technische Universit\"at Darmstadt and Karlsruhe Institute of Technology, Otto-Berndt-Str. 3, 64287, Darmstadt, Germany}
\affiliation{Institute of Nanotechnology, Karlsruhe Institute of Technology, Hermann-von-Helmholtz-Platz 1, 76344 Eggenstein-Leopoldshafen, Germany}

\author{Benedikt Eggert}
\affiliation{Faculty of Physics and Center for Nanointegration Duisburg-Essen (CENIDE), University of Duisburg-Essen, Lotharstr. 1, 47057 Duisburg, Germany}

\author{Leonardo Velasco}
\affiliation{Institute of Nanotechnology, Karlsruhe Institute of Technology, Hermann-von-Helmholtz-Platz 1, 76344 Eggenstein-Leopoldshafen, Germany}

\author{Xiaoke Mu}
\affiliation{Institute of Nanotechnology, Karlsruhe Institute of Technology, Hermann-von-Helmholtz-Platz 1, 76344 Eggenstein-Leopoldshafen, Germany}

\author{Johanna Lill}
\author{Katharina Ollefs} 
\affiliation{Faculty of Physics and Center for Nanointegration Duisburg-Essen (CENIDE), University of Duisburg-Essen, Lotharstr. 1, 47057 Duisburg, Germany}

\author{Subramshu S. Bhattacharya}
\affiliation{Nano Functional Material Technology Centre (NFMTC), Department of Metallurgical and Materials Engineering, Indian Institute of Technology Madras, 600036 Chennai, India}

\author{Heiko Wende}
\affiliation{Faculty of Physics and Center for Nanointegration Duisburg-Essen (CENIDE), University of Duisburg-Essen, Lotharstr. 1, 47057 Duisburg, Germany}

\author{Robert Kruk}
\affiliation{Institute of Nanotechnology, Karlsruhe Institute of Technology, Hermann-von-Helmholtz-Platz 1, 76344 Eggenstein-Leopoldshafen, Germany}

\author{Richard A. Brand}
\affiliation{Institute of Nanotechnology, Karlsruhe Institute of Technology, Hermann-von-Helmholtz-Platz 1, 76344 Eggenstein-Leopoldshafen, Germany}
\affiliation{Faculty of Physics and Center for Nanointegration Duisburg-Essen (CENIDE), University of Duisburg-Essen, Lotharstr. 1, 47057 Duisburg, Germany}

\author{Horst Hahn}\email{horst.hahn@kit.edu}
\affiliation{Joint Research Laboratory Nanomaterials – Technische Universit\"at Darmstadt and Karlsruhe Institute of Technology, Otto-Berndt-Str. 3, 64287, Darmstadt, Germany}
\affiliation{Institute of Nanotechnology, Karlsruhe Institute of Technology, Hermann-von-Helmholtz-Platz 1, 76344 Eggenstein-Leopoldshafen, Germany}


%

\begin{abstract}
High entropy oxides (HEOs) are single phase solid solutions consisting of 5 or more cations in approximately equiatomic proportions. In this study, we show reversible control of optical properties in a rare-earth (RE) based HEO-(Ce\textsubscript{0.2}La\textsubscript{0.2}Pr\textsubscript{0.2}Sm\textsubscript{0.2}Y\textsubscript{0.2})O\textsubscript{2-{$\delta$}} and subsequently utilize a combination of spectroscopic techniques to derive the features of the electronic band structure underpinning the observed optical phenomena. Heat treatment of the HEO under vacuum atmosphere followed by reheat-treatment in air results in a reversible change of the band gap energy, from 1.9~eV to 2.5~eV. The finding is consistent with the reversible changes in the oxidation state and related \textit{f}-orbital occupancy of Pr. However, no pertinent changes in the phase composition or crystal structure is observed upon the vacuum heat treatment. Further annealing of this HEO under H$_2$ atmosphere, followed by reheat-treatment in air, results in even larger but still reversible change of the band gap energy from 1.9~eV to 3.2~eV. This is accompanied by a disorder-order type crystal structure transition and changes in the O~2$p$-RE~5$d$ hybridization evidenced from X-ray absorption near edge spectra (XANES). The O~$K$ and RE~${M_{4,5}}$/$L_{3}$ XANES indicate that the presence of Ce and Pr (in 3+/4+) state leads to the formation of intermediate 4$f$ energy levels between the O~2$p$ and RE~5$d$ gap in HEO. It is concluded that heat treatment under reducing/oxidizing atmospheres affects these intermediate levels, thus, offering the possibility to tune the band gap energy in HEO.

\end{abstract}
\maketitle

\section{Introduction}
High entropy oxides (HEOs) were first reported in 2015\cite{Rost2015} and since then the topic has gained significant interest, which is evident from the numerous published reports focusing on different aspects of HEOs\cite{Dragoe2019, Sarkar2019, Sarkar2018c, Dabrowa2018, Gild2018, Musico2019, Qiu2019, Jimenez-Segura2019,Braun2018}. HEOs can be broadly defined as single-phase solid solution oxides containing 5 or more cations in near-equiatomic compositions. The presence of multiple cations in comparable amounts leads to an enhanced configurational entropy of mixing ($S_{config}$), which is calculated using the Boltzmann statistical entropy equation.\cite{Rost2015,Sarkar2019, Wang2019a} The term \enquote{high entropy oxide} is typically used when the $S_{config}$ of a given oxide system is above 1.5 R.\cite{Gild2018,Sarkar2019,Dabrowa2018,Oses2020} One of the most intriguing characteristics of HEOs is the phase purity despite their compositional complexity. Interestingly, the underlying principle for phase purity is distinct in different HEOs. For instance, in some systems a dominant role of entropy has been evidenced, which effectively overcomes the related enthalpic penalties and stabilizes a single phase solid solution at high temperatures.\cite{Rost2015, Chen2018, Spiridigliozzi2020} 

However, in the majority of HEOs a pertinent role of configurational entropy in stabilizing the phase composition and structure has not been observed, whereas factors such as the oxidation states of specific cations and lattice strain effects govern their phase composition.\cite{Djenadic2017, Dabrowa2018,Gild2018,Sarkar2019,Cheng2019pressure} 
Irrespective of the phase stability mechanisms, the high entropy based design approach in oxides offers an extended compositional flexibility along with retention of the phase purity. Hence, compositions close to the central regions of multi-component oxide phase diagrams can be studied where interesting synergies can be anticipated. A wide range of crystal structures and functional features exhibited by HEOs can be linked to their unique composition and related synergies. \cite{Berardan2016c,Gild2018,Musico2019,Witte2019b,Chen2018b,Sarkar2018c,Qiu2019,Braun2018,Jimenez-Segura2019,Zhang2019,Meisenheimer2017,Wang2019Cat,Sharma2020, Patel2020} Nevertheless, disentangling the role of individual elements is one of the major challenges in HEOs. 

In our previous studies, we have explored the possibility to synthesize several single phase fluorite type HEOs (F-HEOs) by populating the cationic sublattice with 5 or more rare earth (RE) elements\cite{Djenadic2017, Sarkar2017e}. By comparing these compositions, we have observed that the presence of Pr in F-HEOs results in a lowering of the band gap energy. Hence, we speculated that Pr due to its mixed 3+/4+ oxidation state in F-HEOs forms an intermediate unoccupied energy state, which facilitates lower energy electronic transitions of $\sim$2~eV.\cite{Sarkar2017e} Understanding the electronic band structure of conventional binary rare-earth oxides is still a subject of both theoretical as well as practical interest. \cite{Gillen2013, Minasian2017, Petit2005} However, the theoretical band structure calculations for complex F-HEOs are not straightforward. Hence, in this current study, we use different spectroscopic techniques to investigate the validity of our hypothesized energy band diagram in a representative F-HEO composition,  (Ce\textsubscript{0.2}La\textsubscript{0.2}Pr\textsubscript{0.2}Sm\textsubscript{0.2}Y\textsubscript{0.2})O\textsubscript{2-{$\delta$}}. Initially, by means of heat treatment under reducing (vacuum/hydrogen) or oxidizing atmosphere, we have observed reversible changes in band gap energies. The changes in the oxidation states of the RE elements and related variation in the \textit{f} and \textit{d} electronic shell occupancy are studied using X-ray absorption spectroscopy (XAS) and electron energy loss spectroscopy (EELS), which primarily indicates the role of internal redox reactions in determining the electronic characteristics of F-HEO. Although, this kind of redox reaction driven band gap tuning mechanism is known for simple metal oxides,\cite{Ahn2012, Chen2013t} such a possibility has not yet been explored for HEOs. Furthermore, the accompanying physico-chemical changes such as large concentrations of oxygen vacancies, reversible disorder-order transition and Raman features upon H$_2$-treatment make the F-HEO different. Hence, this study not only elucidates the dynamic band structure of F-HEO but also provides a deeper structural insight of these compostionally complex systems.

\section{Experimental}
\subsection{Synthesis}
 F-HEO, (Ce$_{0.2}$La$_{0.2}$Pr$_{0.2}$Sm$_{0.2}$Y$_{0.2}$)O$_{2-\delta}$, was synthesized from water-based precursor solutions of the corresponding rare earth nitrates using reverse co-precipitation techniques (RCP). The precursor solution was added to a basic ammonia solution (pH = 10) in a controlled fashion along with continuous stirring of the mixture at ambient conditions. The formed precipitates were dried at 120~$^\circ$C and further calcined at 750~$^\circ$C in air to achieve the desired crystallographic phase. The temperature, 750~$^\circ$C, has been determined based on the reports of F-HEO synthesized by the nebulized spray pyrolysis  method \cite{Djenadic2017}. 
 The powders obtained after this first 750~$^\circ$C heat-treatment step  in air are defined as the as-synthesised samples. The as-synthesised (or \enquote{as-syn}) samples were subsequently heat treated under different atmospheres, where the temperature, dwell time and the ramping rates were 750 $^\circ$C, 2 hours and 10 $^\circ$C/min, respectively. The heat-treatment conditions were as follows:

\begin{itemize}
    \item Mildly reducing atmosphere: Heat treatment in vacuum (VHT), corresponding to 10$^{-7}$ mbar pressure. 

    \item Highly reducing atmosphere: Heat treatment with flow of 5 \% hydrogen-argon mixture (H$_2$-HT), with a gas flow rate of 50 standard liters per minute. 

    \item Re-heat treatment under air: This was carried out after the VHT or H$_2$-HT, using the respective VHT/H$_2$-HT powder. Hence,  
    the samples obtained are termed as RHT (meaning \enquote{re-heat treated} in air).

\end{itemize}

\subsection{Structural and electronic characterization}

\begin{enumerate}

\item \textbf{X-Ray diffraction (XRD)}: Room temperature XRD patterns were recorded using a Bruker D8 Advance diffractometer with Bragg-Brentano geometry equipped with an X-ray tube having a Cu anode and a Ni filter. Rietveld analysis of the XRD patterns was done using TOPAS V.5.0\cite{Topas2015}. The instrumental intensity distributions were determined using a reference scan of LaB$_6$ (NIST 660a). 

\item \textbf{High resolution transmission electron microscopy (HR-TEM) and Electron energy loss spectroscopy (EELS)}: Specimens for TEM were prepared by directly dispersing the finely ground powders onto a standard carbon coated copper grid. A FEI Titan 80-300 aberration (imaging Cs) corrected transmission electron microscope equipped with a Gatan Tridiem 863 image filter operated at 300 kV was used to examine the specimens. EELS spectra were collected in TEM mode (objective aperture 30 $\mu$m, convergence semi-angle of $<$0.5 mrad, with a dispersion of 0.1~eV/channel, a collection semi-angle of 15 mrad, and an acquisition time of 5 s) and in scanning TEM mode (condenser aperture 70 $\mu$m, convergence semi-angle 14 mrad, dispersion 0.1~eV/channel, collection semi-angle 16 mrad, and acquisition time of 5 s).

\item \textbf{Ultraviolet - visible (UV-Vis) spectroscopy} The UV-Vis spectra were recorded in diffuse reflectance mode in the range from 200 nm to 1200 nm using a Perkin-Elmer Lambda 900 spectrophotometer. From the obtained spectra, optical band gaps were determined by applying the Tauc relation\cite{Tauc1966}: 

$$\left[ F(R_\infty)h\nu\right]^{1/n} =   A(h\nu - E_g)$$								

\noindent								
where $F(R_{\infty})$ is the Kubelka-Munk function, $h$ is Planck’s constant, $\nu$ is the incident frequency, $A$ is a constant, and $E_g$ is the band gap energy. The exponent $n$ denotes the nature of the optical transitions. In this study, $n = 1/2$ has been used which is valid for for direct allowed transitions. The band gap energy values were calculated from linear regression at the inflection point of the $\left[ F(R_\infty)h\nu\right]^2$ vs. h$\nu$ (Tauc) plots. The obtained $h\nu$-intercept values were taken as the band gap values.

\item \textbf{Raman spectroscopy}: Raman spectra were recorded using a Renishaw Raman microscope using infrared (785 nm) and green (532 nm) laser in the range of 200 – 2000~cm$^{-1}$, with a spot size  $\sim 1 \mu$m and laser power of 0.5 mW. All the spectra were the result of 10 accumulations, each lasting 20 s.

\item \textbf{X-Ray absorption spectroscopy (XAS)}: XAS was utilized to study the valence states and the chemical environment of the different constituents. For the investigation of the RE-4$f$ and the O-2$p$ states, measurements were performed in the soft X-ray regime at the XUV diffractometer endstation, located at the beamline UE46$\_$PGM-1\cite{Weschke2018} (BESSY II, Berlin), while for the investigation of the RE-5$d$ states, measurements in the hard X-ray regime have been performed at the beamline P65 \cite{Welter2019} (PETRA III, Hamburg). Due to the large difference of the used photon energy, different measurements procedures were used. For the measurements in the soft X-ray regime (O-$2p$ and RE-$4f$ states) the sample powder was pressed into Indium foil to ensure electrical grounding, while the signal was measured in TEY. In the hard X-ray regime (RE-$5d$ edges), the sample powder was pressed into a pellet, while the absorption measurements were performed in transmission mode. Because of the presence of many atomic species in these samples, careful subtraction of the background signal was necessary. More fundamentally, because of the computational difficulty of calculating the near-edge spectra from first principles in these complex mixed oxides, we have restricted our analysis here to a description of individual white line changes and  comparisons with well established literature results. A recent review of the possibilities of XANES studies has been given by Henderson et al. \cite{Henderson2014} and in a larger context by Wende \cite{Wende2004c}. 

\end{enumerate}

\section{Results}
The results obtained for heat-treatment of (Ce$_{0.2}$La$_{0.2}$Pr$_{0.2}$Sm$_{0.2}$Y$_{0.2}$)O$_{2-\delta}$-F-HEO in vacuum differ significantly to those obtained when heat-treated in hydrogen atmosphere. Hence, two sub-sections specific to the heat-treating atmosphere are made. 
\subsection{Vacuum heat treatment}

\subsubsection{Changes in the optical properties upon VHT}
\begin{figure}[h!]
\[\includegraphics[width=0.85\columnwidth]{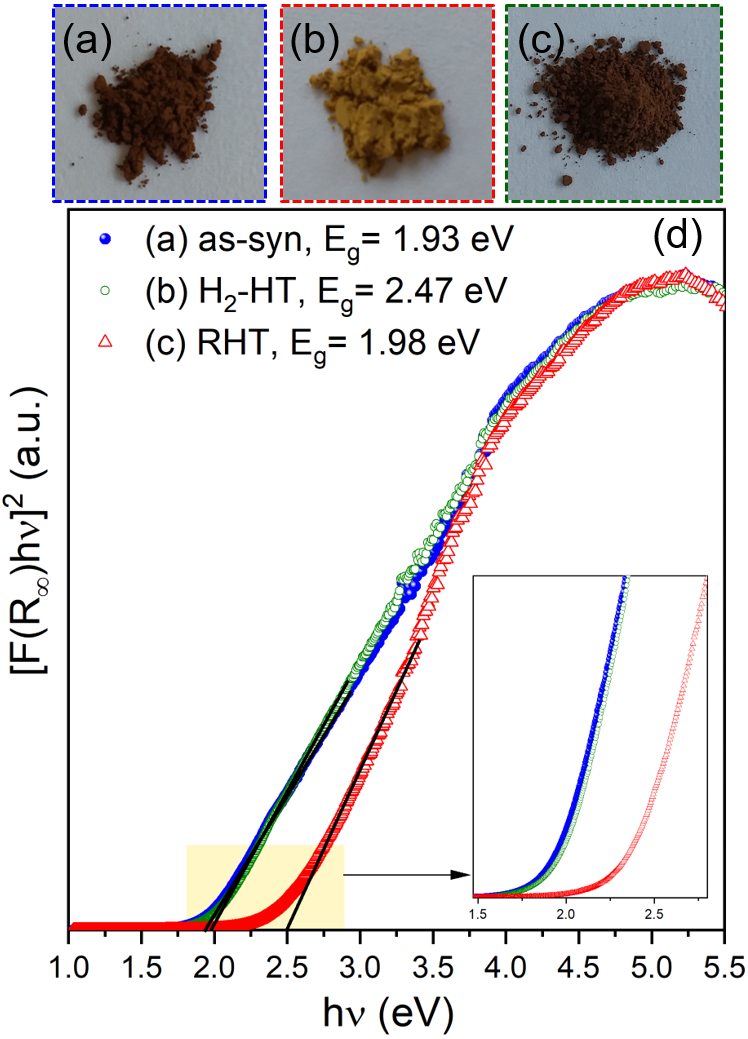}\]
\caption{Photographs of (Ce$_{0.2}$La$_{0.2}$Pr$_{0.2}$Sm$_{0.2}$Y$_{0.2}$)O$_{2-\delta}$ powder, where (a) is the as-synthesized powder, (b) VHT and (c) RHT after the VHT. (d) Corresponding Tauc plot obtained from the UV-Vis spectra of these powders confirm reversible tuning of the band gap. The band gap values are within the 5 \% error bar, i.e, $\pm$0.1~eV}
\label{Fig1}
\end{figure}

As-synthesized F-HEO powder has been  subjected to vacuum heat treatment (VHT) followed by subsequent re-heat treatment in air (RHT). An immediate indication of a reversible change can be observed by simple inspection of the powder color as shown in Figure \ref{Fig1}a-c. The as-synthesized powder is dark brown in color (Fig.~\ref{Fig1}a), whereas VHT results in a yellow colored powder (Fig.~\ref{Fig1}b) and upon further RHT the initial brown color (Fig.~\ref{Fig1}c) is regained. Likewise, the diffuse reflectance spectra of the powders support this observation and the corresponding changes in the band gap energies have been estimated, as presented in Fig.~\ref{Fig1}d. The band gap energy of the as-synthesized powder is 1.93~eV. Upon VHT, an increase in the band gap to 2.47~eV can be observed. The RHT of the powder in air leads to a reversal of the of band gap to 1.98~eV. This value is close to the band gap of the as-synthesized system and the difference ($\sim$5 \%) falls within the experimental error of the measurement. 

\subsubsection{Structural analysis and phase composition upon VHT}

\begin{figure}[h!]
\[\includegraphics[width=1\columnwidth]{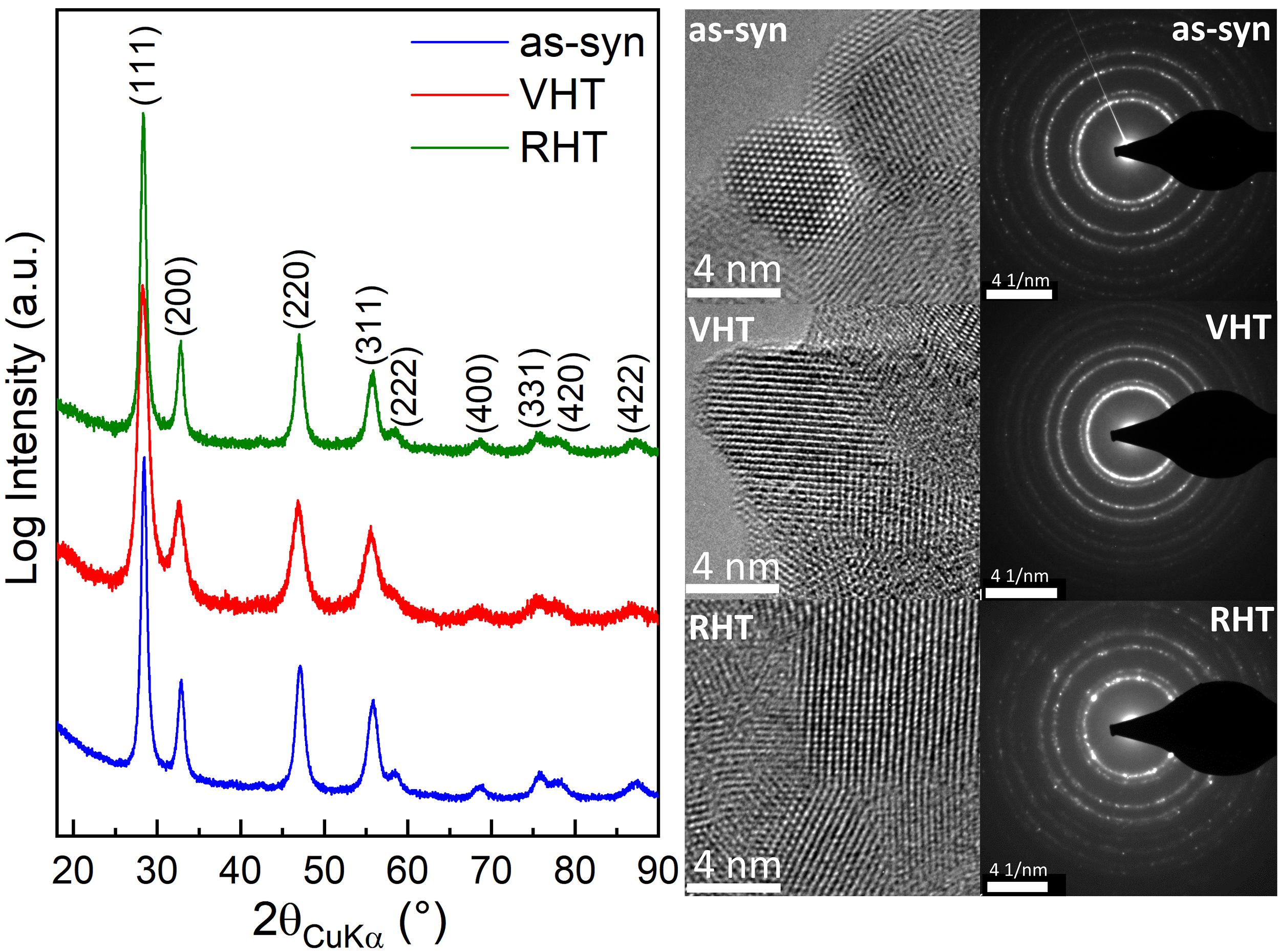}\]
\caption{XRD patterns confirming the phase purity of as-synthesized, vacuum heat treated (VHT) and air re-heat treated (RHT) F-HEO. The TEM micrographs are in good agreement with the XRD results confirming the nanocrystallinity and phase purity at the local level.}
\label{Fig2}
\end{figure}

One of the main characteristics of HEOs is their phase-purity, hence it is important to investigate the effect of heat treatment on the phase composition of F-HEO. Fig.~\ref{Fig2} presents the XRD patterns, HR-TEM micrographs and selected area electron diffraction (SAED) patterns of the as-synthesized, VHT and RHT samples. HR-TEM micrographs along with the XRD patterns reveal the presence of nano sized crystallites, which aggregate to form micron sized particles. Importantly, no change in the crystal structure or phase composition can be 
observed upon heat treatment under varying atmospheres. Structural details obtained from Rietveld refinement of the XRD patterns are presented in Supplemental Figure SI1. A marginal increase  in the lattice parameter ($\sim$0.4 \%) along with broadening of the peaks due to strain effects can be observed upon VHT (Fig.~\ref{Fig2}). These observations hint towards the reduction of some of the constituent cations (resulting in larger ionic sizes) and related oxygen vacancies formation, which is the focus of the following section.

\subsubsection{Oxidation state and charge compensation upon VHT}

In our previous report,\cite{Sarkar2017e} we inferred that the band gap energy of $\sim$2~eV in F-HEO is closely associated with the presence of multivalent Pr$^{3+}$/Pr$^{4+}$. Following this speculation it can be expected that the change/reduction in the oxidation state of Pr will affect the occupancy of the Pr~4$f$ states.  Hence, a change in the band gap energy of F-HEOs can be expected. In order to investigate the change in the oxidation of constituent elements upon VHT, EELS measurements have been carried out.

\begin{figure}[h!]
\[\includegraphics[width=0.95\columnwidth]{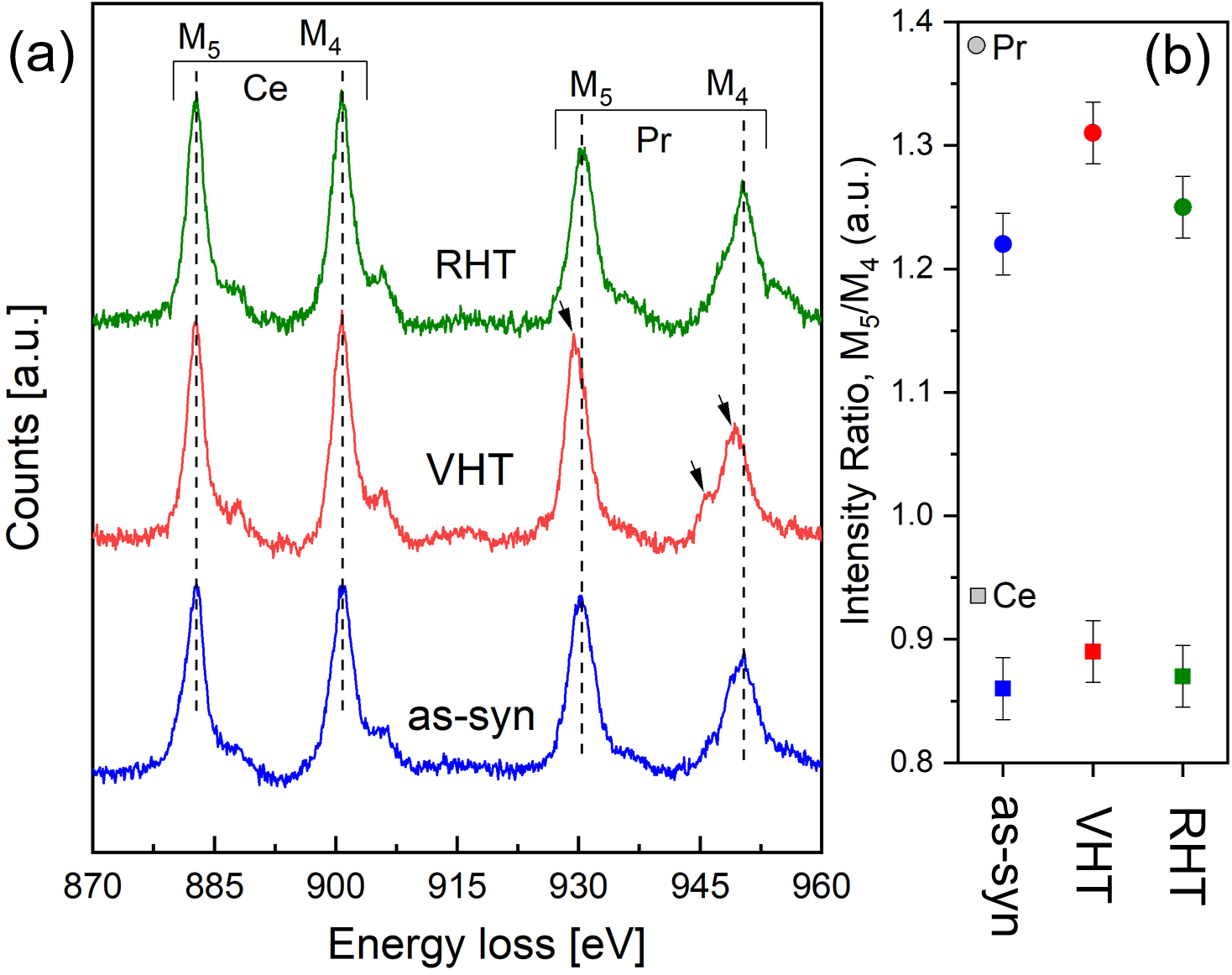}\]
\caption{(a) EELS spectra of the F-HEO showing the Ce and Pr $M_{4,5}$ edges. The chemical shift of Pr edges along with the shoulder at lower energy losses to $M_4$ indicates of the reduction of Pr upon VHT. (b) Average Ce and Pr I$_{M_5}$/I$_{M_4}$ intensity ratios calculated from (a) indicates the \textit{f}-shell occupancy.}
\label{Fig3}
\end{figure}

EELS measurements have been performed on the as-synthesized, VHT and RHT powders. The spectra for Ce and Pr~$M_{4,5}$ edges are displayed in Fig.~\ref{Fig3}. The $M_{4,5}$ whitelines in EELS spectra of the rare-earths (REs) result from electronic transitions between the initial state $3d$  to the $4f$ orbitals, such as $3d_{3/2} \rightarrow 4$f$_{5/2}$ (\textit{M$_4$}) and $3d_{5/2} \rightarrow 4$f$_{7/2}$ (\textit{M$_5$}). The relative intensities ratio (I$_{M_5}$/I$_{M_4}$) and chemical shift of the \textit{M$_{4,5}$} lines denote the change in valency (or the $f$-shell occupancy) of the respective RE cations.\cite{DAngelo2016}. The EELS spectra for Ce and Pr are of utmost importance as these two elements exhibit the highest tendency to vary between the 3+ and 4+ states whereas La, Sm and Y are known for their stable 3+ electronic configuration. For Ce, two whitelines at 882 and 902~eV are observed, which are related to \textit{M$_5$} and \textit{M$_4$} transitions, respectively.\cite{DAngelo2016,Ou2008} However, no noticeable chemical shifts of these lines have been observed after VHT or RHT. This finding is further supported by the fact that the I$_{M_5}$/I$_{M_4}$ ratio is relatively invariant with respect to the heat treatment (Fig.~\ref{Fig3}b). Unlike Ce, the scenario is different for Pr. For the as-synthesized system, Pr \textit{M$_{4,5}$} whitelines are observed at 931 and 951~eV, respectively. A significant chemical shift (of $\sim$0.6~eV) towards low energies occurred upon VHT which indicates a reduction of Pr from a pronounced 4+ state to 3+ dominated multivalent (3+/4+) state \cite{Bowman2016}. The formation of Pr$^{3+}$ is supported by the pronounced shoulder at the left (or lower energies) of the Pr~\textit{M$_4$} line. Additionally, an increase in I$_{M_5}$/I$_{M_4}$ of Pr is also observed upon VHT, further confirming the decrease of Pr$^{4+}$ (Figure \ref{Fig3}(b)). Likewise, upon subsequent RHT of the VHT sample in air, a reverse behavior, i.e., an oxidation of Pr towards the initial Pr$^{4+}$ dominated mixed valence state could be observed.

The reduction of Pr with VHT should be compensated internally. One way of charge compensation can be via oxidation of the other constituent cations, which is not the case here. Hence, the formation of oxygen vacancies can be expected.  We have utilized  Raman spectroscopy to study this possibility. Two characteristic bands centered at $\sim$452~cm$^{-1}$ and $\sim$580~cm$^{-1}$ in all samples, are used for the estimating of the V$_O$ concentration (see Supplementary Information Figure SI3). The $\sim$452~cm$^{-1}$ band corresponds to the F$_{2g}$ stretching mode of fluorite structure while the latter corresponds to the presence of V$_O$.\cite{Filtschew2016,Weber1993} The integral intensity ratio of V$_O$ and the F$_{2g}$ can be used for a relative estimation of the V$_O$ concentration in different systems.\cite{Guo2011b} The inset in Supplementary Information Figure SI2 shows an increase in the amount of V$_O$ concentration upon VHT, while V$_O$ concentrations in the as-synthesized and the RHT samples are nearly comparable (see related discussion in Supplementary Information Section 3). Although the V$_O$ and the charge state of the elements are interlinked, it should be noted that the band gap energy in F-HEO is rather related to the presence of redox-active Pr and its unoccupied 4$f$ intermediate band. This has been observed in our earlier report as well, where we compared various F-HEO compositions with varying amount of oxygen vacancies.\cite{Sarkar2017e} Nonetheless, we have probed samples using Raman spectroscopy, as the change in the V$_O$ concentration upon heat treatment supports the EELS data and indicates the reversible behavior in F-HEO.

To summarize, we observe  a reversible change in the band gap upon VHT of F-HEO followed by RHT. The increase in the energy gap upon VHT is expected due to the decrease in the number of unoccupied Pr~$4f$ states, as evidenced from the EELS results. The reduction of Pr and resulting increase in V$_O$ also support the increase in the lattice parameter and microstrain upon VHT. Furthermore, the fact which makes F-HEO unique is retention of the single phase fluorite structure, even upon VHT. Stability of a fluorite type oxide structure strongly depends on the presence of a 4+ cation, which in VHT-F-HEO is only Ce and a minor part of Pr. Likewise, the V$_O$ concentration (as shown from Raman spectroscopy) in VHT-(Ce$_{0.2}$La$_{0.2}$Pr$_{0.2}$Sm$_{0.2}$Y$_{0.2}$)O$_{2-\delta}$ is rather high compared to conventional binary or doped fluorite type oxides, and the $\delta$ is expected to be $\sim$0.4. To further investigate the impact of stronger reducing atmospheres on physico-chemical features of F-HEO heat treatment under hydrogen atmosphere has been carried out. 

\subsection{Heat treatment under highly reducing H$_2$ atmosphere}

\begin{figure}[h!]
\[\includegraphics[width=0.7\columnwidth]{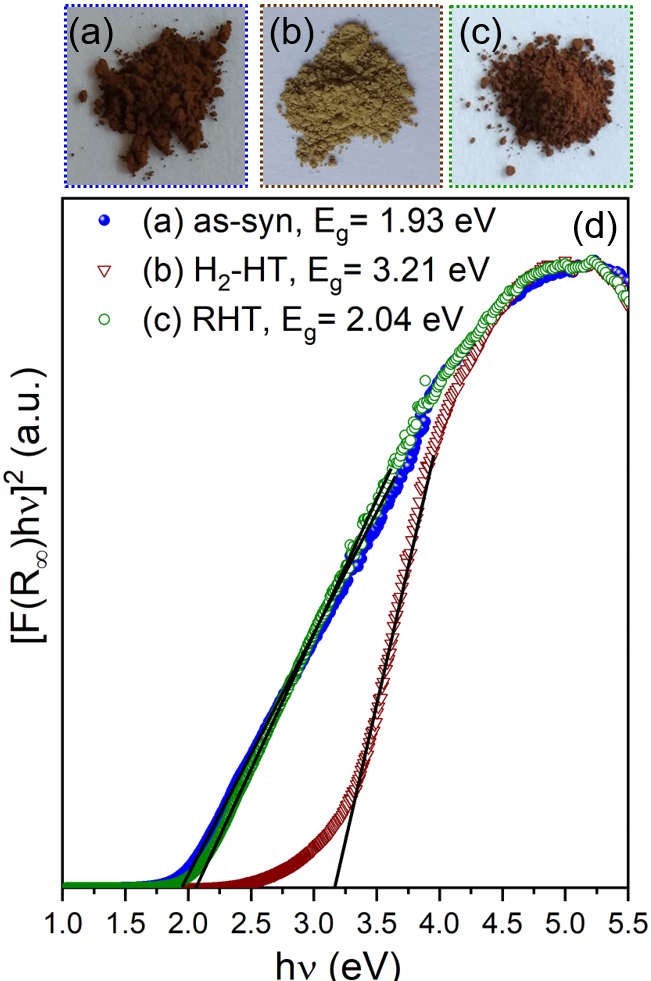}\]
\caption{Photographs and Tauc plot obtained from the UV-Vis spectra of (Ce$_{0.2}$La$_{0.2}$Pr$_{0.2}$Sm$_{0.2}$Y$_{0.2}$)O$_{2-\delta}$ powder, where (a), (b) and (c) corresponds to the as-synthesized, hydrogen heat treated (H$_2$-HT) and re-heat treated samples, respectively. The band gap values are within the 5~\% error bar, i.e, $\pm$0.1~eV.}
\label{Fig4}
\end{figure}

(Ce$_{0.2}$La$_{0.2}$Pr$_{0.2}$Sm$_{0.2}$Y$_{0.2}$)O$_{2-\delta}$ has been heat treated  under a highly reducing 5~\%~H$_2$ + Ar atmosphere (H$_2$-HT), followed by re-heat treatment in air (RHT). A reversible change in the powder color can be observed after this treatment, as is shown in Figure~\ref{Fig4}a-c. Importantly, the change in the band gap energy is larger than what was observed for VHT, i.e., increasing from 1.93~eV to 3.21~eV (Fig.~\ref{Fig4}d). RHT of the H$_2$-HT powder in air leads to reversion of the band gap to 2.04~eV. This band gap tailoring of around 60 \% ($\sim$ 1.3~eV) can be considered as significant, especially given that the change is reversible. 

\subsubsection{Structural analysis and phase composition upon H$_2$-HT}

\begin{figure}[htb]
\[\includegraphics[width=1\columnwidth]{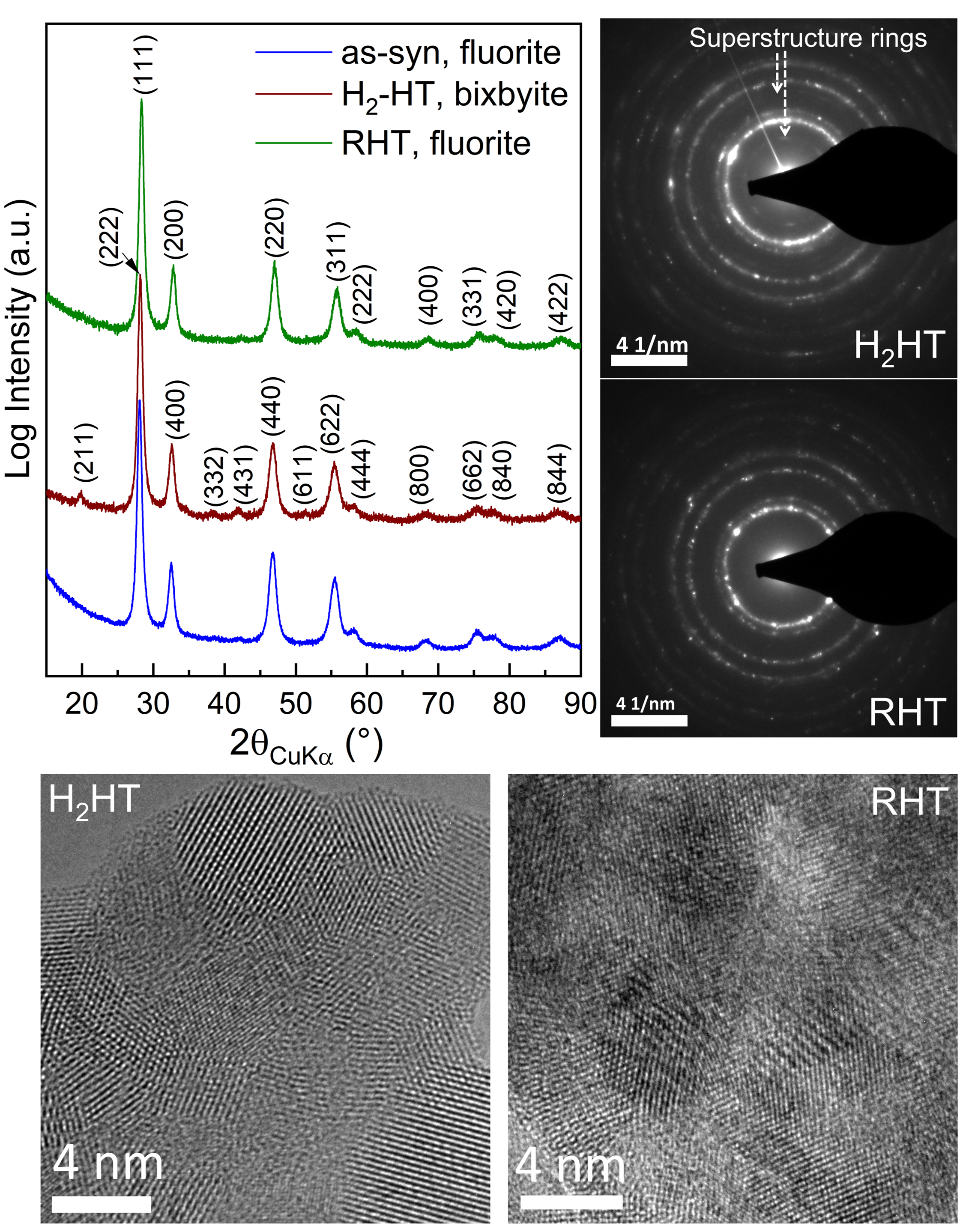}\]
\caption{A reversible transition from fluorite to a bixbyite structure is observed upon hydrogen heat treatment (H$_2$-HT) followed by air re-heat treatment of (Ce$_{0.2}$La$_{0.2}$Pr$_{0.2}$Sm$_{0.2}$Y$_{0.2}$)O$_{2-\delta}$. The TEM micrographs and SAED patterns confirm the reversible disorder-order type structural transition.}
\label{Fig5}
\end{figure}

Fig.~\ref{Fig5} presents the XRD patterns along with SAED micrographs for the H$_2$-HT and the RHT samples. Rietveld refinements of the XRD patterns are provided in Supplemental Figure SI3. Unlike for  the as-synthesized and RHT samples, we observe additional peaks in the XRD patterns for the H$_2$-HT sample. These additional peaks are the superstructure reflections indicating a structural transition from fluorite ($Fm\bar3m$) to a C-type bixbyite ($Ia\bar3$) structure. Evidence of faint superstructure reflections can also be observed in the SAED pattern (Figure \ref{Fig5}) which further confirms the formation of a single phase bixbyite structure upon H$_2$-HT. 

Bixbyite is a well-known low symmetry oxygen deficient body centered cubic variant of the fluorite family, where one out of every four oxygen ions is missing.\cite{Chiang1997, Ou2006} The symmetry relationship between fluorite ($a$) and bixbyite ($2a$) includes a twofold increase in the lattice parameter. Thus for direct structural comparisons, the pseudosymmetric lattice parameter of bixbyite structure (i.e., half of the actual lattice parameter) is considered. In fact the as-synthesized (Ce$_{0.2}$La$_{0.2}$Pr$_{0.2}$Sm$_{0.2}$Y$_{0.2}$)O$_{2-\delta}$ sample undergoes a similar symmetry lowering when heat treated in normal atmospheric conditions (i.e., under air)\cite{Djenadic2017,Sarkar2017e}. However, there are two subtle differences between the symmetry lowering happening upon normal heat treatment in air (as observed in our earlier studies) and the H$_2$-HT. First of all, upon H$_2$-HT, this phase transition happens at a much lower temperature, i.e., already at 750 $^\circ$C, whereas the phase transition will happen only above 1000 $^\circ$C if heat treated under normal air atmosphere\cite{Djenadic2017,Sarkar2017e}. Secondly, and a rather unusual phenomenon, is that the phase transition to bixbyite upon H$_2$-HT is reversible, i.e., a single phase fluorite structure can be regained when the H$_2$-heat treated sample is re-heated (RHT) in air at 750 $^\circ$C (see XRD and SAED in Fig.~\ref{Fig5}). However, we observed\cite{Sarkar2017e, Djenadic2017} that the transition to a bixbyite phase is completely irreversible if the as-synthesized sample is heat treated above 1000 $^\circ$C in normal air atmosphere.\cite{Djenadic2017,Sarkar2017e} As mentioned above, the transition from fluorite to bixbyite is mostly governed by the presence of oxygen vacancies which start to order above a certain temperature. One of the reasons for the structural transition already at 750 $^\circ$C upon H$_2$-HT can be the formation of 
additional oxygen vacancies due to stronger reducing atmosphere. This is supported by increase ($\sim$ 0.5\%) in the pseudosymmetric lattice parameter upon H$_2$-HT. In fact, 750 $^\circ$C under H$_2$ might already be sufficient to order the oxygen vacancies. However, the temperature might still not be large enough to result in a stable ordering, making the transition back to fluorite upon RHT possible. Further discussion of this point is presented in the section C.

\begin{figure}[h!]
\[\includegraphics[width=1\columnwidth]{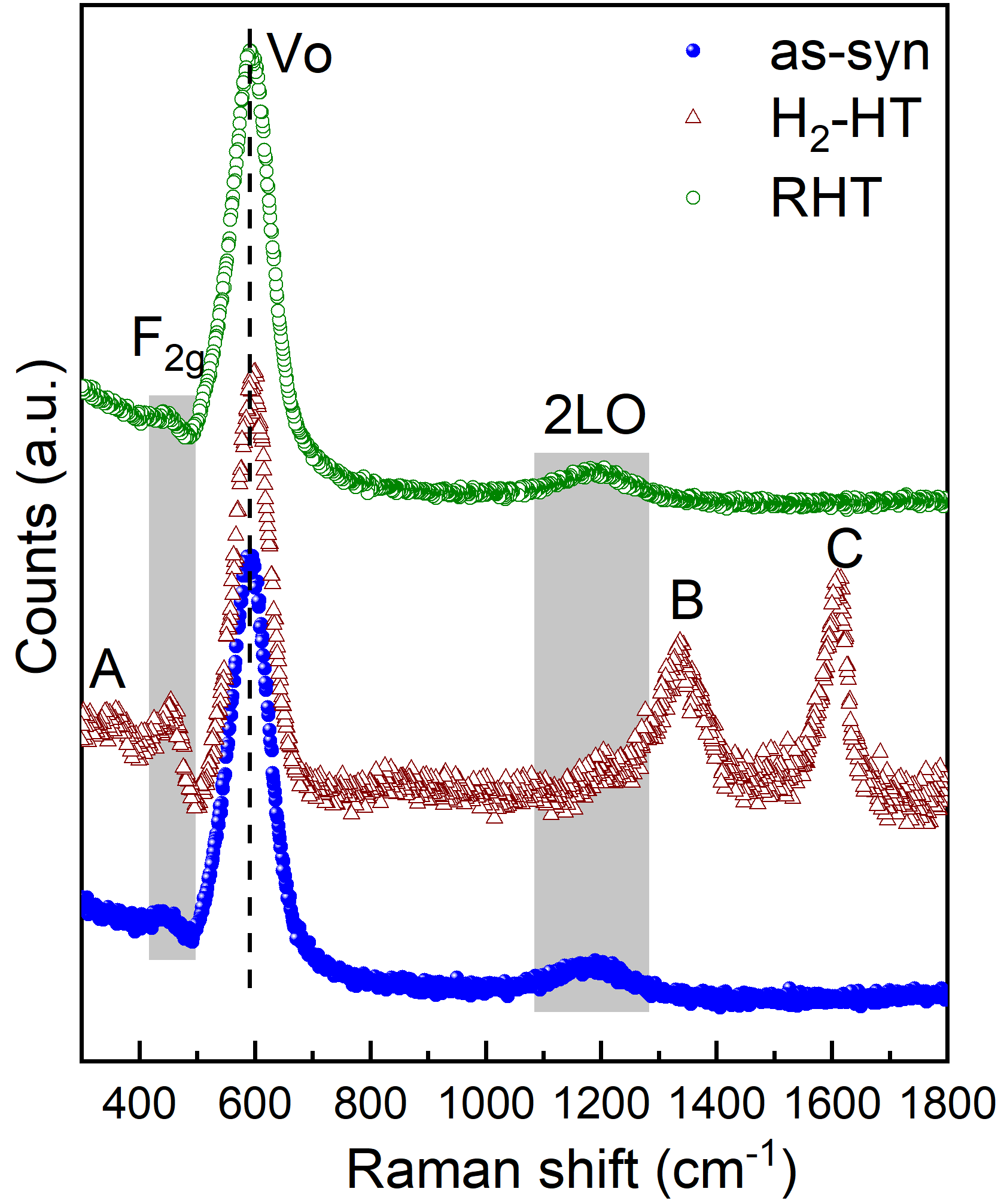}\]
\caption{Raman spectra of the (Ce$_{0.2}$La$_{0.2}$Pr$_{0.2}$Sm$_{0.2}$Y$_{0.2}$)O$_{2-\delta}$ using 532 nm excitation laser. Three additional bands centered at 358~cm$^{-1}$ (A), 1333~cm$^{-1}$ (B) and 1610~cm$^{-1}$ (C) are observed in the H$_2$-HT system.}
\label{Fig6}
\end{figure}

We have used  Raman spectroscopy to further investigate these systems due to its sensitivity to local chemical changes, especially related to light scatterers such as oxygen. It should be noted that the infrared laser which we typically use to study F-HEO could not be used for the H$_2$-HT sample due to a possible resonance effects (see Supplementary Information Figure SI4). Hence, a green laser ($\lambda$ = 532 nm) has been used for comparison. It can be seen from Figure~\ref{Fig6}, that  all the F-HEO systems show distinct Raman bands at $\sim$452~cm$^{-1}$, $\sim$1180~cm$^{-1}$ and $\sim$580~cm$^{-1}$. By comparing these with the parent binary (or doped) rare earth based oxides (containing Ce/La/Pr/Sm/Y), 
we can draw correlations with CeO$_2$ based systems \cite{Schilling2017, Filtschew2016, Dilawar2008, Schmitt2020, Guo2011b}. The band at $\sim$452~cm$^{-1}$ can be associated with the triply degenerated F$_{2g}$ vibration mode\cite{Weber1993}. The band at $\sim$1180~cm$^{-1}$ can be related to the second order 2\textit{LO} Raman mode, which is generally observed in ceria or its doped variants.\cite{Filtschew2016, Weber1993} The strong broad band $\sim$580~cm$^{-1}$ can be best fitted using a two band model, which can be attributed to the presence of oxygen vacancies ( V$_O$) and its related bonding with the 3+ and 4+ cations\cite{Schilling2017}. Given the fact that bixbyite (as in H$_2$-HT) is a superstructure of fluorite, additional vibrations bands, or strengthening/rupturing of some existing ones can be expected. After H$_2$-HT, three additional bands appear. The band $\sim$358~cm$^{-1}$ (A in Fig.~\ref{Fig6}) can be attributed to a second order Raman mode arising from a combination of A$_{1g}$, E$_{g}$, and F$_{2g}$ scattering tensors \cite{Filtschew2016}. However, we do not find a proper match for the two prominent additional bands at $\sim$1333~cm$^{-1}$ and $\sim$1610~cm$^{-1}$ which are observed only upon H$_2$-HT (see Fig.~\ref{Fig6}). Furthermore, these bands cannot be directly attributed to the crystal structural change, as the bixbyite variant of Ce$_{0.2}$La$_{0.2}$Pr$_{0.2}$Sm$_{0.2}$Y$_{0.2}$)O$_{2-\delta}$ obtained upon normal air heat-treatment does not exhibit these features (see Raman spectra in Supplementary Information Figure SI4). Interestingly, both the bands disappear upon subsequent RHT of the H$_2$-HT sample in air and a spectrum similar to the as-synthesized system is obtained. One of the plausible reasons for appearance of the additional bands upon H$_2$-HT can be due to possible hydrogen incorporation in the F-HEO system. Hydrogen being a light element can lead to strong vibrational modes. Recently, it has been reported that H incorporation (even in hydride form) is possible in (bulk) fluorite type CeO$_2$ when it is heat treated above 350 $^\circ$C in H$_2$ atmosphere\cite{Wu2017}. Such a H-incorporation cannot be ignored in F-HEO as H$_2$-HT is done at 750 $^\circ$C. Moreover, the presence of large  V$_O$ concentrations in F-HEO might potentially facilitate H-incorporation. However, this is a speculation and finding concrete evidence of the presence of H is not simple. Hence, experiments sensitive to H, such as inelastic neutron scattering, are needed to support such a hypothesis.

In order to further understand the other physico-chemical changes happening upon H$_2$-HT, XAS has been performed.

\subsection{Element resolved charge state and electronic structure of F-HEO}\label{sec:exp_xanes}

XAS probes the (unoccupied) electronic density of states above the Fermi energy in an element specific way\cite{Sthr1992,Wende2004c}. Hence, it offers the possibility to disentangle the element specific contributions in complex systems like F-HEO. Each of the XAS spectra have been normalised to the edge jump. In the following, the respective changes at the different absorption edges are discussed. In addition, an investigation concerning changes of the amount of unoccupied states has been performed.\cite{Thole1992,Carra1993} 

\subsubsection{Rare Earth $M_{4,5}$ edges: 3d$\rightarrow$ 4$f$ and $L_3$ edge: 2$p$ $\rightarrow$ 5d}
\begin{figure}[htb]
\[\includegraphics[width=1\columnwidth]{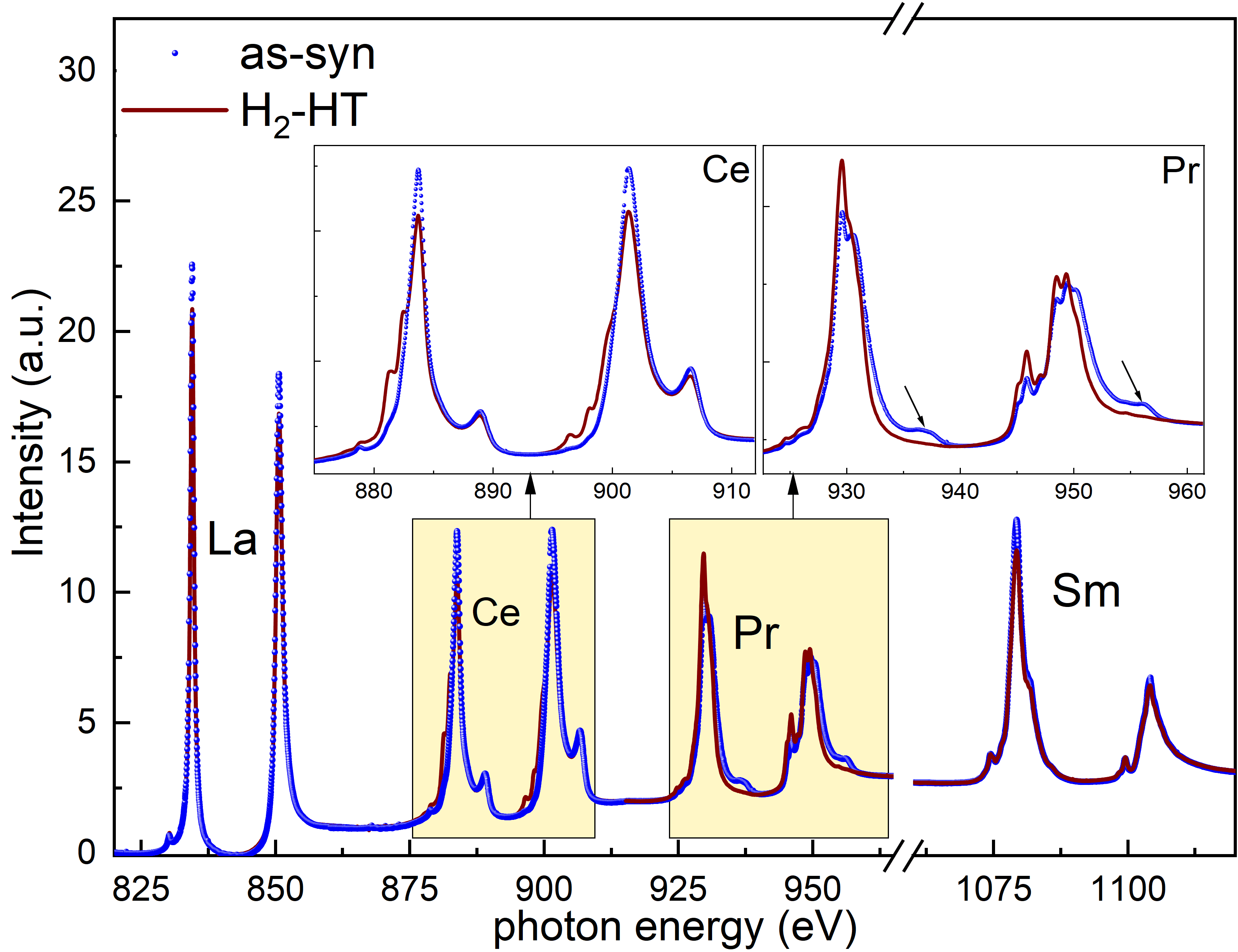}\]
\caption{XAS spectrum at the different RE $M_{4,5}$ edges for the as-synthesized and H$_2$-heat treated sample. The respective spectrum have been normalised to the edge jump after the $M_4$ edge. The insets show a magnification of the Ce and Pr edges.}
\label{REM45}
\end{figure}

At the $M_{4,5}$-edges the transition from the 3$d$ states into the unoccupied 4$f$ states occurs (in a dipole approximation). The respective absorption spectra for the as-synthesized and H$_2$-HT sample are shown in Fig.~\ref{REM45}. The La $M_{4,5}$ edges indicate that both the as-synthesized and H$_2$-HT samples are in 3+ valence state, while a reduction of the free states by $\sim$3.5\,\% can be determined by the difference of the integrated spectral area. We see a similar, intuitive, behavior for the Sm which is also in a 3+ valence state in both cases, while the integrated spectral area is reduced by 3.9\,\% upon H$_2$-HT. For Ce a detailed fine structure occurs prior to both the $M_5$ and $M_4$ edges after H$_2$-HT (inset Fig.~\ref{REM45}). These additional peaks can be assigned to the Ce$^{3+}$, while the initial peaks can be assigned to Ce$^{4+}$.\cite{Alayoglu2013,Heyraud2013} Therefore, it can be said that Ce is present in a 4+ dominated multivalent (3+/4+) state after H$_2$-HT. Due to the occurring multivalency in Ce, a change of the number of the unoccupied states can only be determined from the integrated spectrum. From this analysis, an increase of 0.9\,\% of the integrated whiteline intensity can be determined - such an increase can be neglected and can be explained by differences in the normalisation procedure. Pr on the other hand is present in a multivalent 3+/4+ state (with a pronounced 4+ nature), as indicated by the prominent after edges\cite{Herrero2011} in the as-synthesized sample (see inset of Fig.~\ref{REM45}). It changes into a (fully) 3+ state upon H$_2$-HT. Furthermore, the unoccupied 4$f$ states in Pr are reduced by $\sim$15\%.

Further, the RE $L_3$ edges are also studied as the hard X-rays in transmission mode provides a better idea about the bulk nature of the sample. In the $L_3$ edge, the transition from the 2$p_{3/2}$ states to unoccupied 5$d$ states occurs. Ce and Pr $L_3$ spectra are shown in Figure \ref{REL3}. A complete spectrum including La and Sm is provided in the Supplementary Information Figure SI5. For the $L_3$ absorption edge, the discussion is simpler compared to the $M_{4,5}$ edges. For La and Sm a 3+ state is evident, while no significant change of the whiteline intensity is visible. For Ce, 4+ state can be observed in the as-synthesized sample, while reduction to a mixed valent state is evident upon H$_2$-HT. The presence of Ce$^{3+}$ results in the shoulder A\cite{Schmitt2020}, as shown in the inset of Fig.~\ref{REL3}. Nevertheless, a pronounced Ce$^{4+}$ state is maintained even upon H$_2$ with no major change in the whiteline intensity. For Pr a significant change of the spectral structure is visible. The as-synthesized sample shows a distinct multiplet spectrum (4$f^2$ and 4$f^1$ in inset Fig.~\ref{REL3}). From the literature it is known, that such a spectrum is typically observed in multivalent Pr$^{3+/4+}$, where the Pr$^{4+}$ is ascribed to the 4$f^1$ peak\cite{Garc2011, Ogier2019}. After H$_2$-HT the change of the spectral near edge fine structure indicates a complete reduction to Pr$^{3+}$. Accompanied with this change is a reduction of the spectral area by $\sim3\,\%$. This can be explained either by a change of the unoccupied 5d states or structural changes due to the EXAFS signal from the La L$_2$ edge ($E_{0,\mathrm{La L}_2} =5891$~eV). For this a detailed EXAFS modelling would be necessary, which is difficult due to the relative small available $k$-space for this complex system as visible in Supplementary Information Figure SI5.

\begin{figure}[htb]
\[\includegraphics[width=0.8\columnwidth]{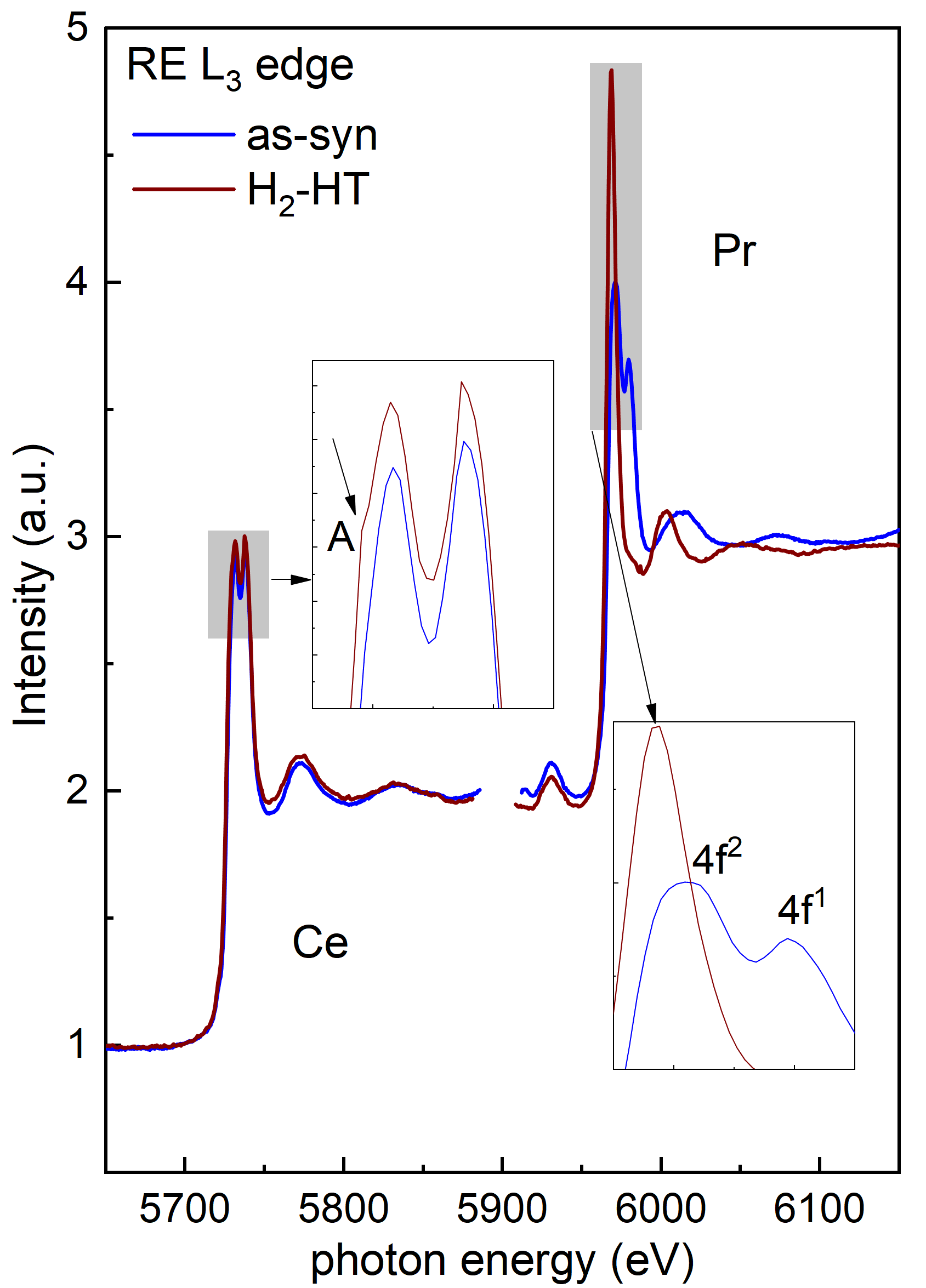}\]
\caption{XAS spectra at the Ce and Pr $L_3$ edges for the as-synthesized and H$_2$-heat treated sample. The peak for $E=5930$~eV originates from the La L$_2$ edge.}
\label{REL3}
\end{figure}

\subsubsection{Oxygen $K$-edge: 1$s$~$\rightarrow$ 2$p$}

At last, we focus on the hybridization of the O~2$p$ states with the RE neighbors by probing the O~1$s\rightarrow$2$p$ transition at the O~$K$ edge.  Unlike the RE cations, the XANES spectrum of O-$K$ edge is rather complex, as O hybridizes with all the RE cations. Hence, disentangling all the spectral features is not straightforward, given the fact that XANES of F-HEO (or similar complex systems) have not been reported on before. Nevertheless, the changes happening in the RE $M_{4,5}$ and $L_3$ edges help us to identify the major spectral variation in O-$K$ edges upon heat treatment. The prominent features are observed at the photon energy $E$~=~528.6~eV(A); 530~eV (B); 532~eV (C); 533~eV (D), 537~eV (E) and 539~eV (F), respectively (see Fig.~\ref{OxyK}). Feature A can be attributed to the Pr~4$f$-O~2$p$ hybridization\cite{Minasian2017} (discussing pertaining to this feature is explained in the following Section IV). The features B, D and E have been attributed to O~2$p$-states, which hybridize with the 4$f$, 5d-e$_g$ and 5d-t$_{2g}$ levels of Ce in stoichiometric CeO$_2$, respectively\cite{Minasian2017, Rodriguez2003}. The splitting of the 5$d-e_g$ and 5$d-t_{2g}$ states occurs only if crystal field effects lift the degeneracy\cite{Soldatov1994,Rodriguez2003},  indicating that the changes in spectral features D and E are due to an alteration of the local structure (evident from Fig.~\ref{Fig5}). The relative increase of the peak intensity of feature D compared to feature E reveals a local ordering of the oxygen vacancies. This finding is in agreement with reports on doped CeO$_2$ \cite{Ou2006, Ou2008} and supports the formation of C-type bixbyite structure upon H$_2$-HT. Most likely the mild reduction of Ce$^{4+}$ upon H$_2$-HT, along with the increased oxygen vacancies stemming strong reduction of the Pr (as well as Ce), can be the reason for the formation of the bixbyite structure. The feature F (539~eV) can be matched with the O-2$p$ and Sm-5$d$ hybridization\cite{Chen2014, Altman2016}, hence do not change upon H$_2$-HT. Likewise, the presence of feature C can also be observed both in the as-synthesized as well as the H$_2$-HT system which is possibly linked to the O~2$p$-RE~5$d$ hybridization.

\section{Discussion}

\begin{figure}[h!]
\[\includegraphics[width=1\columnwidth]{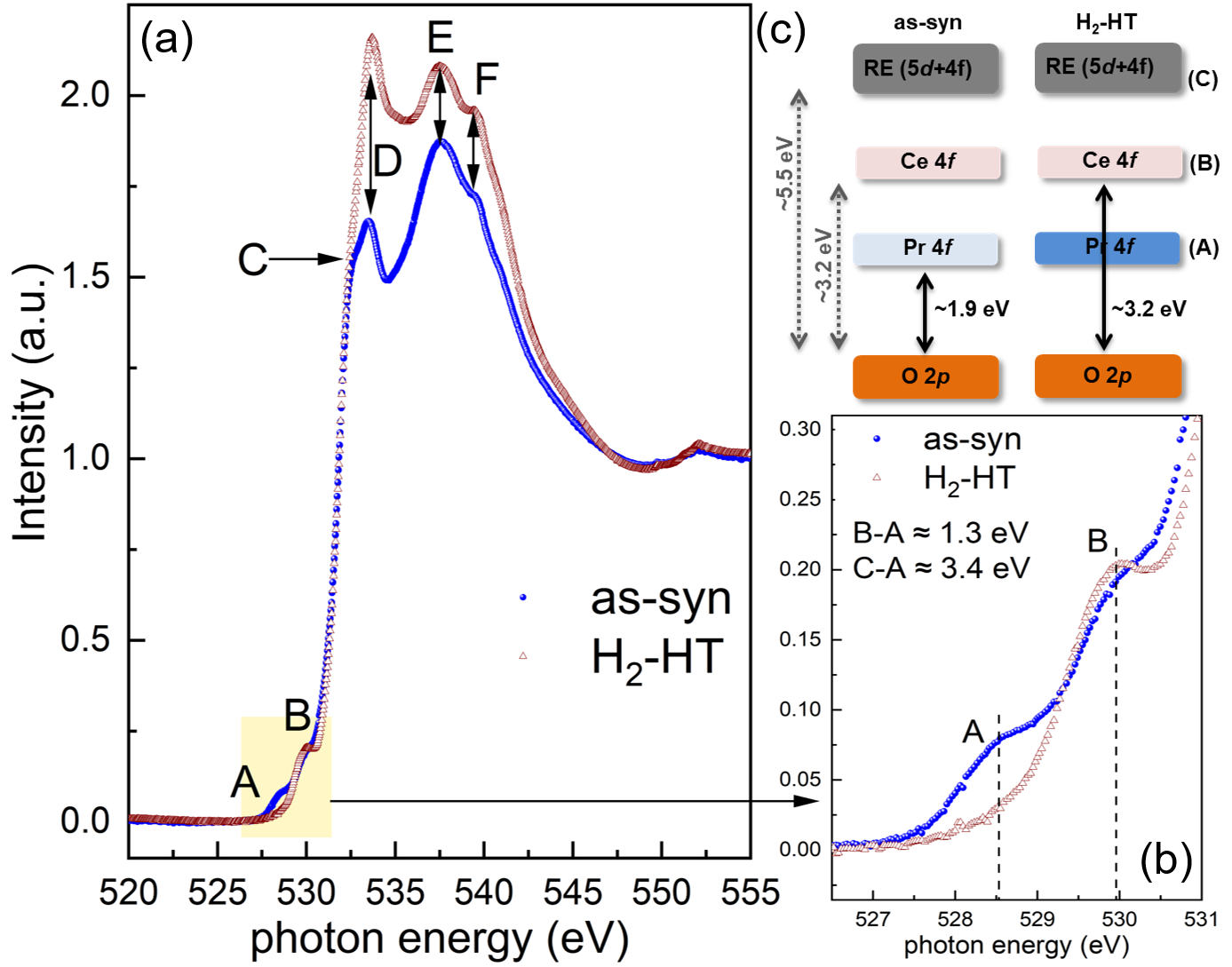}\]
\caption{(a) O $K$ edge X-ray absorption spectroscopy spectra of the as-synthesised and H$_2$ heat treated sample. The spectra are normalised to the edge jump. (b) Magnification of the pre edge region.  (c) Schematic of the electronic band diagram of as-synthesized (oxidized) and H$2$-heat treated (reduced) F-HEO}
\label{OxyK}
\end{figure}

In conventional rare earth (RE) oxides, the occupied oxygen 2$p$ states are hybridized with RE 5$d$/4$f$ states, resulting in the formation of the bonding and the anti-bonding orbitals. The O~2$p$ with the occupied RE 5$d$/4$f$ state forms the valence band, which is generally classified as the O~2$p$ band. While the anti-bonding states originating from the RE-O hybridization are the conduction bands, which are generally denoted in terms of RE state related to the specific hybridization, like RE~5$d$/4$f$. Thus, the electronic transition from the O~2$p$ state to the unoccupied RE~5$d$/4$f$ states determines the band gap energy. In fact the changes in the electronic structure of binary rare earth oxides can be directly correlated to the changes observed in O-$K$ edges\cite{Altman2016}. Likewise, in the case of F-HEO, the analysis of the O-$K$ XANES edge provides a comprehensive understanding of the electronic band structure and valance states (which is also supported by the other spectroscopic techniques). The most interesting features in O-$K$ edge of F-HEOs are A and B. As mentioned above feature B corresponds to the Ce~4$f$-O~2$p$ hybridization, while feature A corresponds to Pr~4$f$-O~2$p$ hybridization. Feature B is present in both cases, however, the intensity of the feature A decreased to almost zero after H$_2$-HT. This indicates distinct changes in Pr~4$f$ states which are related to the reduction in Pr valency or decrease in  the unoccupied states in the 4$f$ level (see  Pr-$M_{4,5}$ edges Fig.~\ref{REM45}). Here, we would like to mention that the band gap energy difference (see Fig.~\ref{Fig4}) between the as-synthesized and the H$_2$-HT is $\sim$1.3~eV. This value matches almost perfectly the energy difference between feature A (Pr~4$f$-O~2$p$, 528.6~eV) hybridization and feature B (Ce-4$f$-O-2$p$, 529.9~eV). The weakening of the feature A means increase in the occupation of the Pr-4$f$ band, thus prohibiting electronic transitions to that same. This finding validates our band energy diagram (see Fig.~\ref{OxyK}c), where we expect that the electronic transitions will occur from the O-2$p$ (valence bands) to Pr-4$f$ (unoccupied) state in case of the as-synthesized system. However, upon reduction of Pr the 4$f$ level will get filled and hence, the transition to unoccupied Ce-4$f$ will determine the band gap. This is exactly what we observed in the O-$K$ edge (see Fig.~\ref{OxyK}c), when the as-synthesized and the H$_2$-HT systems are compared. The fact that feature B becomes prominent upon H$_2$-HT, even upon the reduction of Ce$^{4+}$ can be explained as follows. Firstly, the reduction in Ce$^{4+}$ to Ce$^{3+}$ is rather mild, as the predominance of Ce$^{4+}$ can be observed from the Ce~$M_{4,5}$/$L_3$ edges (Fig.~\ref{REM45} and \ref{REL3}). Additionally, no major changes in the whiteline intensity of Ce~$M_{4,5}$ indicates only negligible changes in Ce-4$f$ occupancy.

Furthermore, as can be seen in Fig.~\ref{OxyK}a, both the Pr~4$f$-O~2$p$ and Ce~4$f$-O~2$p$ are actually pre-edge features, while the main absorption corresponding to the RE~5$d$-O~2$p$ hybridization starts above 531~eV. This is equivalent to our hypothesized energy band diagram ( Fig.~\ref{OxyK}c) with the two intermediate energy bands at lower energy gaps, while the gap between O2$p$ to unoccupied RE-5$d$ is much larger. The energy difference between feature A and C is $\sim$3.5~eV (so is the energy difference between the onset of edge A to edge C). This means the effective gap between O-2$p$ and RE-5$d$ is $\sim$5.5~eV. In fact this value is in good agreement with our earlier hypothesis\cite{Sarkar2017e} and is close to the expected energy gap in rare earth oxides without intermediate 4$f$ states.\cite{Gillen2013, Altman2016} Hence, it can be summarized that in F-HEO Ce and Pr lead the formation of intermediate energy states resulting in narrow band gaps, whereas the other 3+ cations add to main energy gap $\sim$5.5~eV. Depending upon the heat treatment condition, i.e. reducing/oxidizing atmospheres, each of these electronic structure in F-HEO can be achieved reversibly.

\section{Conclusions}

This work attempts to provide a comprehensive understanding of the tunable electronic band structure in a fluorite type rare-earth based high entropy oxide (F-HEO). Owing to the element specific techniques used here, it is possible to disentangle the individual effects of the constituent cations, which can be directly correlated to the observed reversible changes in the optical features. Reversible changes ($\sim$0.5 eV) in the band gap energies are achieved in F-HEO upon vacuum heat treatment followed by reheating in air. Importantly, the single phase fluorite type structure of (Ce\textsubscript{0.2}La\textsubscript{0.2}Pr\textsubscript{0.2}Sm\textsubscript{0.2}Y\textsubscript{0.2})O\textsubscript{2-{$\delta$}} is maintained even upon vacuum heat treatment with Ce being the sole element which is purely 4+. This shows the potential of F-HEO to accommodate large amount of oxygen vacancies. Stronger reduction of F-HEO, achieved via hydrogen heat treatment, allows for larger change ($\sim$1.2 eV) of the band gap energy, which can be reverted back to the initial state upon subsequent annealing in air. In addition, a reversible structural transition from fluorite to a single phase oxygen vacancy ordered C-type bixbyite structure has been observed. Raman spectroscopy further strengthens this reversible behavior, as local chemical features before and after reheat treatment are identical. X-ray absorption near edge spectra (XANES), using soft and hard X-rays, indicate that the crystal structure transition can be correlated to the reduction of Ce$^{4+}$, while tuning of the band gap energy is related to the reduction of Pr and related changes in Pr~4$f$-O~2$p$ hybridization.

Although practical applications using F-HEO are yet to materialize, the unique combination of narrow and tunable band gap energies along with large concentrations of oxygen vacancies in F-HEO might be beneficial for applications related to mixed electronic-ionic conduction, catalysis etc. An electrochemical approach to tune the band gap in F-HEO can be another prospective research direction. In any case, further experiments and most importantly support from theoretical studies are essential to exploit complete potential of F-HEO.

\begin{acknowledgements}

A.S. and H.H. acknowledge financial support from the Deutsche Forschungsgemeinschaft (DFG) project HA 1344/43-1. B.E. and H.W. acknowledge financial support from the Deutsche Forschungsgemeinschaft (DFG) project WE2623/14-1. X.M. and L.V. acknowledge the support of Karlsruhe Nano Micro Facility (KNMF) for the access to the TEM. A.S. thanks Soumabha Bag (INT, KIT) for fruitful discussions.  We thank the Helmholtz-Zentrum Berlin for the allocation of beamtime at the beamline UE46 PGM${_1}$ (proposal 192-08578-ST/R) and Eugen Weschke for the support during the beamtime. We acknowledge DESY (Hamburg, Germany), a member of the Helmholtz Association, for the provision of experimental facilities. Parts of this research were carried out at PETRA III (proposal 20190485) and we would like to thank Ruidy Nemausat for assistance in using beamline P65.

\end{acknowledgements}

\section*{Supplemental information}
The supplemental information includes the Rietveld refinements of the XRD patterns, Raman spectra (of vacuum heat treated F-HEO and bixbyite HEO) and XANES of La and Sm $L_3$ edges. 

%

\end{document}